\documentclass[letterpaper]{article}

\usepackage[T1]{fontenc}
\usepackage{newtxtext,newtxmath} 

\usepackage{geometry}
\usepackage{setspace}

\usepackage{achemso}

\usepackage{graphicx}
\usepackage{float}
\newfloat{scheme}{htbp}{los}
\floatname{scheme}{Scheme}
\floatname{chart}{Chart}
\newfloat{graph}{htbp}{loh}

\usepackage{endnotes}

\usepackage{authblk}
\author[1]{Michał J. Grzybowski}
\affil[1]{University of Warsaw, Faculty of Physics, Pasteura 5, Warsaw, Poland}
\author[2,3,4]{Katarzyna Gas}
\affil[2]{Center for Science and Innovation in Spintronics, Tohoku University, 2-1-1 Katahira, Aoba-ku, Sendai 980-8577, Japan}
\affil[3]{Laboratory for Nanoelectronics and Spintronics, Research Institute of Electrical Communication, Tohoku University, 2-1-1 Katahira, Aoba-ku, Sendai 980-8577, Japan}
\affil[4]{Institute of Physics, Polish Academy of Sciences, Aleja Lotników 32/46, PL-02668 Warsaw, Poland}
\author[4]{Jaroslaw Z. Domagala}
\author[4]{Serhii Kryvyi}
\author[5]{Łukasz~Gondek}
\affil[5]{Department of Solid State Physics, Faculty of Physics and Applied Computer Science, AGH University of
Kraków, Mickiewicza 30 Avenue, 30-059 Kraków, Poland}
\author[6]{Soumik Das}
\affil[6]{School of Physical Sciences, Indian Association for the Cultivation of Science,
2A \& B Raja S. C. Mullick Road, Jadavpur, Kolkata 700032, India}
\author[1]{Jan Suffczyński}
\author[4]{Oleksii Liubchenko}
\author[4]{Anna Kaleta}
\author[4]{Sławomir~Kret}
\author[4]{Maciej~Sawicki}
\author[1]{Wojciech Pacuski}

\title{High-Volume Wurtzite MnSe in MnSe/CdSe Superlattices: Structure, Magnetism and Optical Response}
\date{*Email: Michal.Grzybowski@fuw.edu.pl}

\begin{document}

\maketitle

\begin{abstract}
Polymorphism in Mn-based chalcogenides offers access to magnetic semiconductors with distinct crystal symmetries, polar structures, and different magnetic orders. In particular, wurtzite MnSe exhibits an appealing interplay of electric polarization and several competing magnetic phases including altermagnetic order. However, stabilizing metastable polymorphs, such as wurtzite MnSe, in experimentally useful volumes remains challenging. Here, we show that a nanoscale-period MnSe/CdSe superlattice stabilizes wurtzite MnSe over micrometer-scale total thickness while suppressing the formation of the thermodynamically favored rock-salt phase. The increased magnetic volume enables bulk-sensitive characterization of this polar polymorph. The room-temperature lattice parameters of wurtzite MnSe are studied by independent experimental techniques and determined to be $a = (4.18 \pm 0.03)\,$\AA\ and $c = (6.78 \pm 0.03)\,$\AA . SQUID magnetometry reveals antiferromagnetic ordering with $T_N = 62\,$K, which is accompanied by an anomaly of the MnSe lattice parameter, as oberved by temperature-dependent X-ray diffraction. Optical measurements reveal excitonic transitions pointing to the band gap at around $3.4\,$eV. These results highlight the utility and importance of heterostructure engineering for detailed studies of metastable phases of magnetic semiconductors, in particular wurtzite MnSe.
\end{abstract}

\section*{Keywords}

manganese selenide, superlattice, antiferromagnets, altermagnets, molecular beam epitaxy, N\'eel temperature



\section{Introduction}

Combining magnetism and semiconductor physics offers promising potential applications. Consequently, the crossroads of these two fields has attracted sustained research interest over recent decades in a variety of contexts. Nowadays, it is antiferromagnetic spintronics and the physics of altermagnetism, two relatively young disciplines, that drive vigorous development of magnetic semiconductor studies, especially devoted to Mn-based chalcogenides. Generally, they aim at utilizing magnetically compensated materials as active elements of electrical devices, for instance in data storage or spin-current generation, motivating an intense search for materials and mechanisms capable of delivering these functionalities.

Collinear antiferromagnets have been attractive and appealing objects of research following the development of methods to control and detect the spin axis reorientation \cite{Wadley2016, Jungwirth2016, Bodnar2018, Baldrati2019, Meer2021}. Non-collinear antiferromagnets have also attracted considerable attention due to the presence of an anomalous Hall effect without net magnetization or spin-currents \cite{Chen2014, Kubler2014, Nakatsuji2015, Kimata2019, Tsai2020, Takeuchi2021}. Finally, a novel classification of collinear magnets has been recently developed, in which altermagnets exhibiting both magnetic compensation and spin-split band structures form a distinct group \cite{Yuan2020, Smejkal2022, Smejkal2022b}. In this context, manganese-based chalcogenides stand out as a rich platform for investigating magnetic phenomena which can host either altermagnetic \cite{Krempasky2024, Hariki2024, Amin2024, Kluczyk2024} or non-collinear phases \cite{Mavani2026, Burlet1997}. As semiconductors they also offer numerous strategies of controlling the strain, electrical and magnetic properties by doping or growth engineering. Their polymorphism \cite{Siol2018, Grzybowski2024, Chen2026, Sadowski2026}, i.e. the ability to adopt different crystalline structures for a single material system, is both an interesting challenge, and a promising avenue for exploration of diverse magnetic structures. Even the same crystal structure seems to have competing, energetically comparable magnetic orders as predicted theoretically \cite{Mavani2026}.

In this paper we focus on wurtzite MnSe epitaxial thin films. Epitaxial layers of wurtzite MnSe have been demonstrated only recently \cite{Grzybowski2024}. The fact that the material may simultaneously host altermagnetic order and a finite electric polarization makes it exceptionally interesting. The electrical polarization is possible in the wurtzite phase - it was predicted theoretically \cite{Bezzerga2025} and inferred experimentally \cite{Grzybowski2026}. Initial theoretical research suggested that altermagnetic phase of MnSe could be the ground state \cite{Grzybowski2024}. More recent calculations showed that different magnetic phases may compete with a non-collinear all-in-all-out spin configuration as a ground state. However, the altermagnetic arrangement of spins should be achievable with small Cr doping \cite{Mavani2026}. The possibility of numerous competing magnetic orders including the altermagnetic one together with an electric polarization make wurtzite MnSe an appealing platform for studying the interplay between the electrical polarity and magnetism.

The experimental data regarding magnetic properties of wurtzite MnSe has been elusive. This is mainly due to the fact that wurtzite is not the most thermodynamically stable phase of MnSe - bulk crystals \cite{Lindsay1951} and epitaxially grown layers on GaAs \cite{Kim1998} show rock-salt symmetry. In the past, superlattices of MnSe/ZnSe were grown by molecular beam epitaxy to stabilize zinc-blende crystal structure \cite{Kolodziejski1986, Chang1987, Klosowski1991, Samarth1991, Heimbrodt1993}. Wurtzite phase was reported in the past as a minor impurity phase \cite{Baroni1938} or in the form of chemically synthesized wurtzite nanoparticles \cite{Sines:2010_AC, Yang2012, Zhang2013, Sarma2014, Yang2017, Sarma2022}. A recently developed growth procedure for epitaxial layers of wurtzite MnSe \cite{Grzybowski2024} allowed structural characterization of the polytype. However, good quality layers were limited to small thicknesses limiting real applications of this material. Here, we present a method to overcome this challenge. We demonstrate for the first time the stabilization of epitaxial wurtzite (wz-) MnSe over a micrometer-scale total thickness by forming a superlattice of wz-MnSe and wz-CdSe. This advance moves wz-MnSe beyond the thin-film, interface-dominated regime into one where bulk-sensitive probes become accessible. It enables the experimental determination of its N\'eel temperature by SQUID magnetometry, and more broadly lays the groundwork for further volume-limited investigations of this metastable phase.

\section{Results and discussion}
\subsection{Epitaxial growth}

\begin{figure}
  \centering
  \begin{minipage}{0.5\textwidth}
  \centering
  \includegraphics[width=0.9\linewidth]{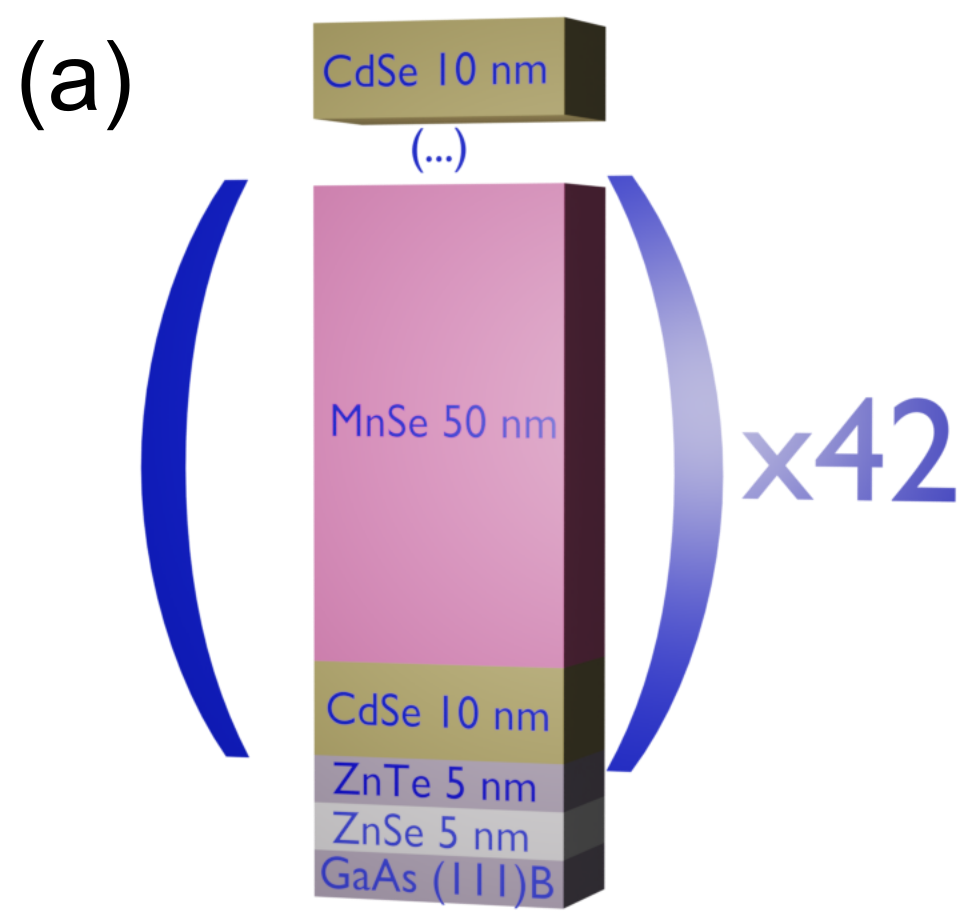}
  \end{minipage}
  \vspace{1cm}
  \begin{minipage}{0.3\textwidth}
  \centering
  \includegraphics[width=1\linewidth]{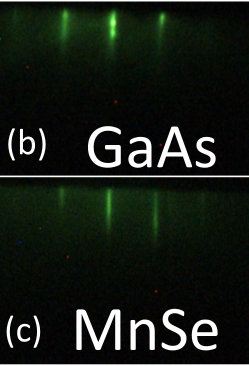}
  \end{minipage}
\caption{\label{fig:mbe} (a) Scheme of the epitaxial wz-MnSe/wz-CdSe superlattice sample. RHEED streak patterns of GaAs (b) substrate just before the growth and (c) of one of the wz-MnSe layers within the considered wz-MnSe/wz-CdSe superlattice.}
\end{figure}

A schematic of the superlattice structure is presented in Fig.~\ref{fig:mbe}(a). It is grown on GaAs (111)B. The substrate is first annealed up to $T=500^{\circ}$C to remove native surface oxides. After a subsequent cooling to $T=200^{\circ}C$, 5-nm-thick ZnSe followed by 5-nm thick ZnTe buffer layers are grown to compensate for the lattice mismatch. Afterwards, the actual superlattice consisting of 42 repetitions of a nominally 50-nm wz-MnSe/10-nm wz-CdSe bilayer is grown. The RHEED streak pattern shown in Fig.~\ref{fig:mbe}(b-c) remains visible throughout the whole growth process. Initially, it is formed of sharp and bright lines. With the increasing thickness of the superlattice the RHEED streak pattern dims and becomes only slightly deformed but was visible until the end of the process. Growth temperature is an important parameter, because elevated growth temperatures promote interdiffusion whereas temperatures that are too low prevent smooth 2D growth of MnSe. Therefore, we focus on this particular sample for optimized growth parameters among the set of differently grown structures.

\subsection{Transmission Electron Microscopy}

\begin{figure}
  \centering
  \includegraphics[width=1\linewidth]{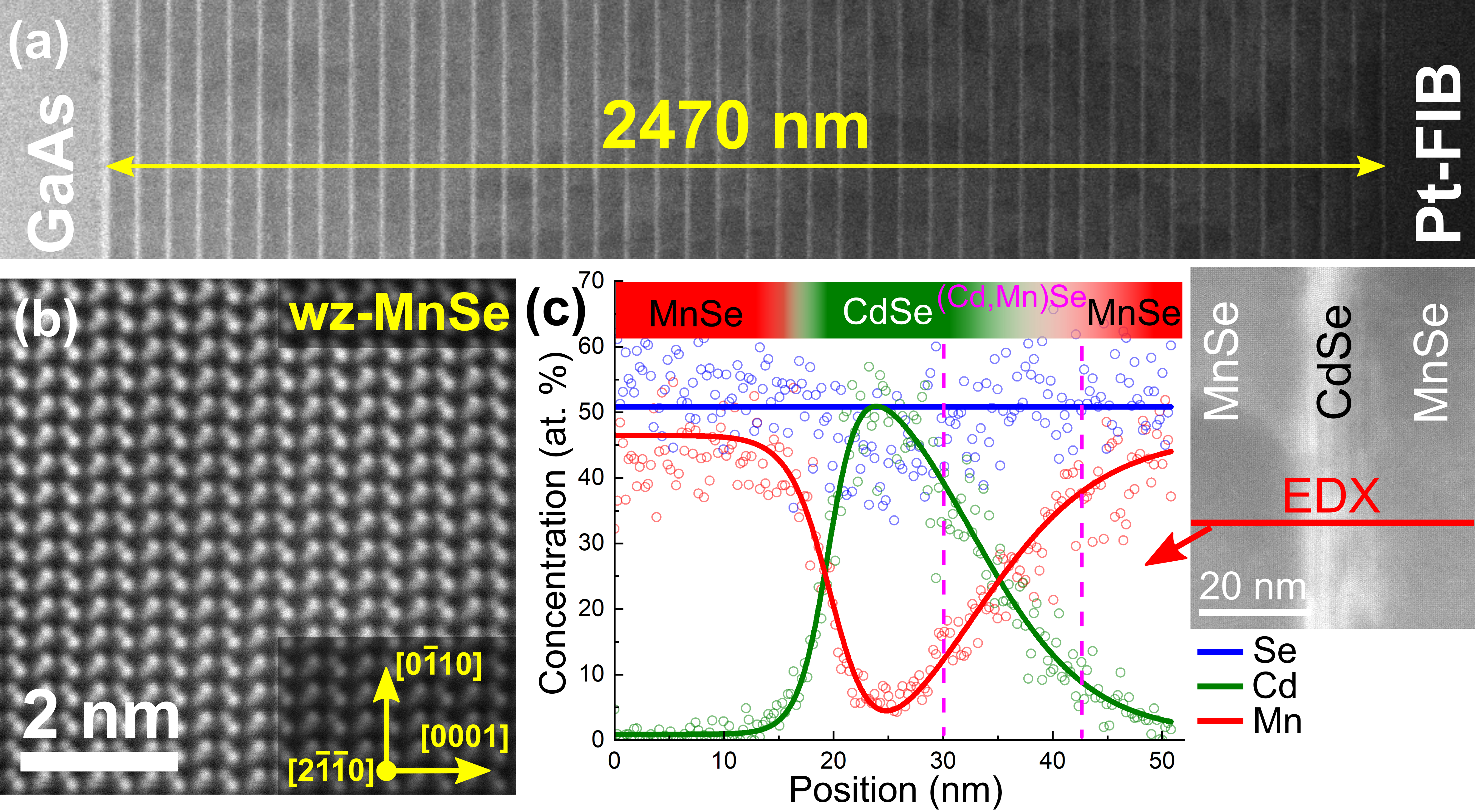}
  \caption{\label{fig:tem} (a) Low-magnification cross-sectional STEM image of wz-MnSe/wz-CdSe superlattice. (b) Atomic-resolution STEM image of a MnSe layer acquired along the $[2\overline{1}\overline{1}0]$ zone axis, showing the wurtzite MnSe structure. (c) STEM-EDX concentration profile across a MnSe/CdSe/MnSe region; open symbols represent the experimental data and solid lines the corresponding fits.}
\end{figure}

The local crystal structure, layer morphology, and interfaces of the superlattice were investigated using transmission electron microscopy (TEM). Cross-sectional scanning transmission electron microscopy (STEM) reveals a well-preserved periodic multilayer structure throughout the entire superlattice thickness of approximately $2.5\,\mu$m (Fig.~\ref{fig:tem}(a)). Individual periods consist of wz-MnSe layers with a representative TEM-measured thickness of approximately 47~nm separated by 10~nm wz-CdSe interlayers, in good agreement with the nominal layer sequence (Fig.~S1). Atomic-resolution STEM imaging of the MnSe layers acquired along $[2\overline{1}\overline{1}0]$ zone axis reveals an atomic-column arrangement consistent with the wurtzite structure (Fig.~\ref{fig:tem}(b)). Additional STEM observations performed near the bottom and the top surface of the superlattice show comparable layer thicknesses and confirm that the wurtzite structure of MnSe is retained across the investigated superlattice thickness (Fig. S1). These observations confirm that the metastable wurtzite polymorph is preserved throughout the multilayer structure.

A representative STEM-EDX profile across a MnSe/CdSe/MnSe region shows complementary variations of the Mn and Cd concentrations, whereas the Se concentration remains approximately constant across the investigated layers (Fig.~\ref{fig:tem}(c)). The elemental distributions reveal a pronounced asymmetry between the two types of interfaces. The MnSe-to-CdSe interface is relatively sharp, whereas the CdSe-to-MnSe interface exhibits a substantially broader compositional transition. At this interface, the Cd concentration gradually decreases over an extended distance while the Mn concentration progressively recovers, resulting in a compositionally graded (Mn,Cd)Se region. The asymmetric interface profiles therefore indicate that intermixing is significantly more pronounced when MnSe is grown on CdSe than when CdSe is grown on MnSe. Such asymmetry was also observed in early studies of II-VI telluride heterostructures  \cite{Gaj:1994_PRB, Grieshaber1994}. A more detailed atomic-resolution analysis of the CdSe interlayers reveals local wurtzite (wz)/ zinc-blende (zb) polytypism (Fig.~S2). While the adjacent MnSe layers retain the wurtzite structure, the CdSe interlayers contain both extended zinc-blende-like regions and shorter local variations of the stacking sequence. The latter can be viewed either as stacking faults within the wurtzite sequence or, equivalently, as ultrathin zinc-blende-like insertions. Twin-related domains are also observed within the zinc-blende-like CdSe regions (Fig.~S2).

The lattice parameters of the wurtzite MnSe phase are additionally determined from high-resolution TEM images. To minimize errors associated with image-scale calibration and residual microscope distortions, the image scale is calibrated using the known lattice parameters of the GaAs substrate, which is assumed to be fully relaxed. The resulting lattice parameters of wurtzite MnSe are $a = (4.18 \pm 0.03)\,$\AA\  and $c = (6.78 \pm 0.03)\,$\AA. 

\subsection{X-Ray Diffraction}

\begin{figure}
  \centering
  \includegraphics[width=0.65\linewidth]{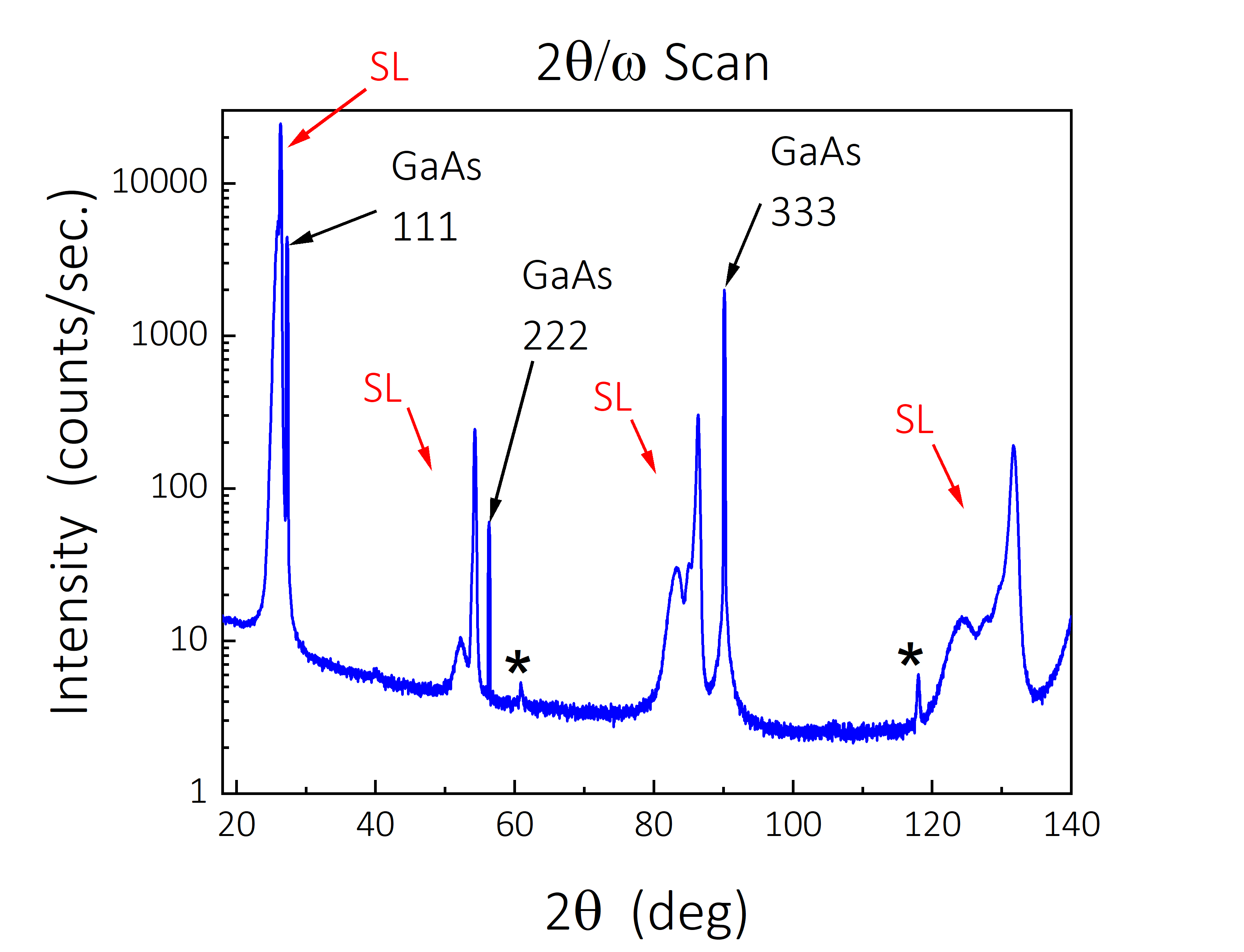}
  \caption{\label{fig:xrd}Low-resolution $2\theta/\omega$ x-ray diffraction scan showing the 111, 222, and 333 reflections from the GaAs substrate, with the wz-MnSe/wz-CdSe superlattice satellites on the lower-angle side of each substrate peak. The sharp, narrow peaks (*) observed at approximately 61$^\circ$ and 118$^\circ$ ($2\theta$) are parasitic, off-specular reflections originating from the high-offcut Si substrate holder due to Z-axis sample displacement. They do not represent any crystalline phases within the deposited SL structure.}
\end{figure}

\begin{figure}
  \centering
  \includegraphics[width=0.65\linewidth]{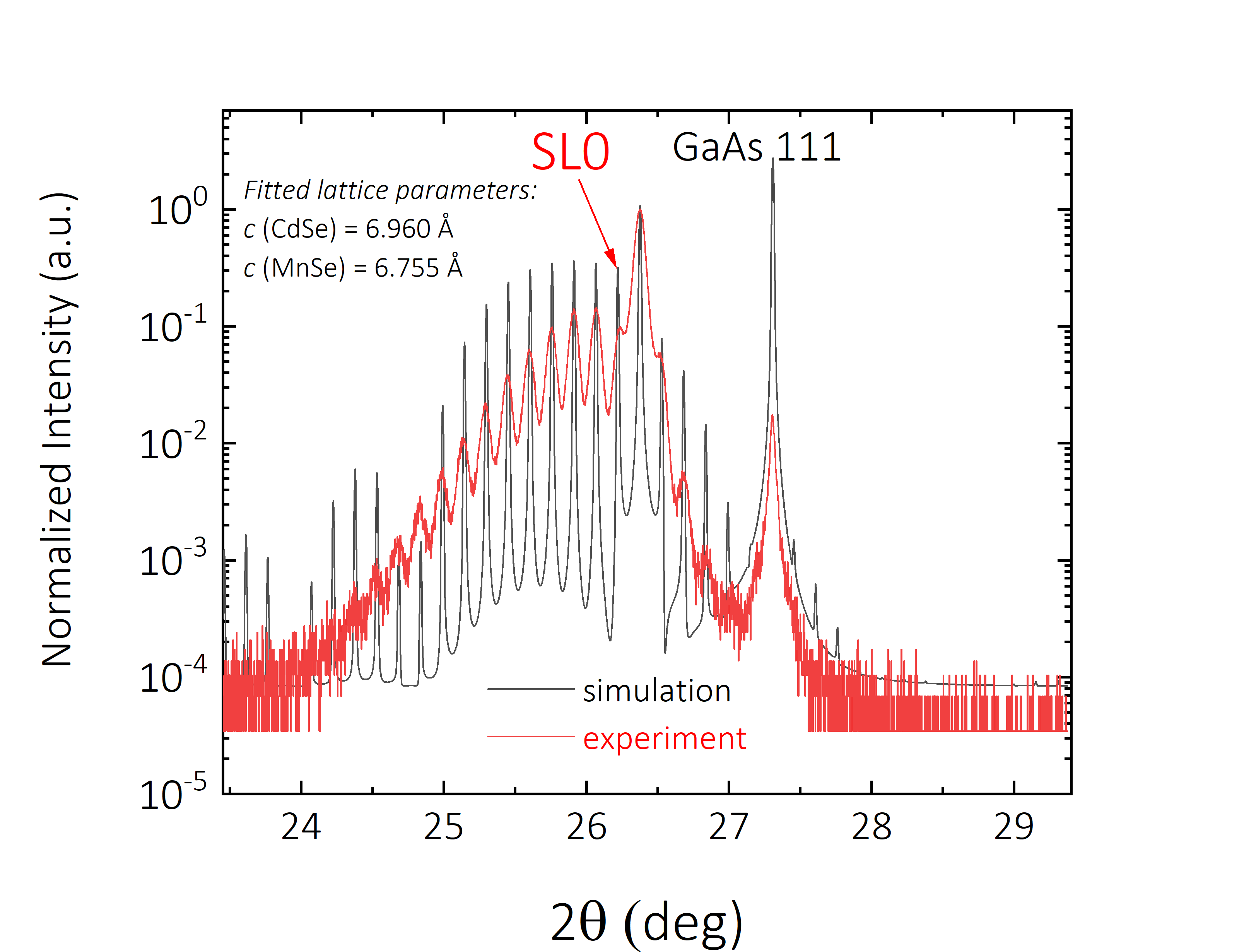}
  \caption{\label{fig:xrdhr}High-resolution $2\theta/\omega$ X-ray diffraction scan in the vicinity of the GaAs 111 reflection compared with a dynamical-diffraction simulation of the 42-period wz-MnSe/wz-CdSe superlattice. The zero-order superlattice reflection (SL0) is indicated. The fitted structural parameters are $c = (6.76 \pm 0.07)\,$\AA\ and $h = (48 \pm 1)\,$nm for MnSe and $c = (6.96 \pm 0.07)\,$\AA\ and $h = 11.2^{+0.7}_{-1.2}\,$nm for CdSe.}
\end{figure}

XRD results are consistent with TEM data. Fig.~\ref{fig:xrd} shows wide-range, low-resolution $2\theta/\omega$ scans of the superlattice (SL) at room temperature: MnSe/CdSe satellite reflections appear on the lower-angle side of each GaAs peak, with the most intense satellite closest to the substrate peak and the zero-order satellite (SL0) substantially weaker (see Fig.~\ref{fig:xrdhr}). Given the large thickness contrast between the layers, this dominant satellite is expected to reflect mainly the MnSe structural response, an assignment supported below and used as a proxy for MnSe in the temperature-dependent analysis. Peak broadening, most pronounced at higher angles, indicates non-negligible structural imperfections. No (002) reflections originating from the rock-salt phase of MnSe are observed \cite{Grzybowski2024}. Such reflections were previously observed in thick epitaxial MnSe layers grown on GaAs, regardless of the substrate orientation. The $\phi$-scans of the asymmetric $(10\overline{1}5)$ reflection (see Supplementary Information) and the TEM results provide complementary evidence supporting the formation of phase-pure wz-MnSe.

A high-resolution $2\theta/\omega$ scan near GaAs (111) (Fig.~\ref{fig:xrdhr}) resolves clear SL satellites, including SL0, compared against a dynamical-diffraction simulation \cite{Zaus1993, Holy2007} of an ideal 42-period wz-MnSe/wz-CdSe superlattice. Differential-evolution fitting of the satellite positions gives the following values of $c$-lattice parameter and layer thickness ($h$): $c = (6.76 \pm 0.07)\,$\AA\ and $h = (48 \pm 1)\,$nm for MnSe as well as  $c = (6.96 \pm 0.07)\,$\AA\ and $h = 11.2^{+0.7}_{-1.2}\,$nm for CdSe. The uncertainties were estimated using the procedure described in Ref.\cite{Bowen:1999_Phil}. The lattice parameters obtained from TEM and XRD agree within their combined experimental uncertainties. Smaller $c$ lattice of MnSe compared with CdSe corresponds to a higher MnSe Bragg angle for individual layers, but in the considered multilayered structure, the observed satellites arise from coherent diffraction by both constituents. Nevertheless, since MnSe accounts for approximately 80\% of the superlattice period, we expect the dominant satellite to be strongly sensitive to the structural changes in MnSe. We therefore use its position as an approximate indicator of the MnSe-related structural response in temperature-dependent XRD. 

The simulated satellite peaks are markedly narrower than the experimental ones, with differences in relative intensities, indicating deviations from the idealized SL structure, consistent with reduced structural quality. This agrees with TEM evidence of compositional grading at the MnSe/CdSe interfaces and zinc-blende inclusions in the wz-CdSe layers.


\begin{figure}
  \centering
  \includegraphics[width=0.75\linewidth]{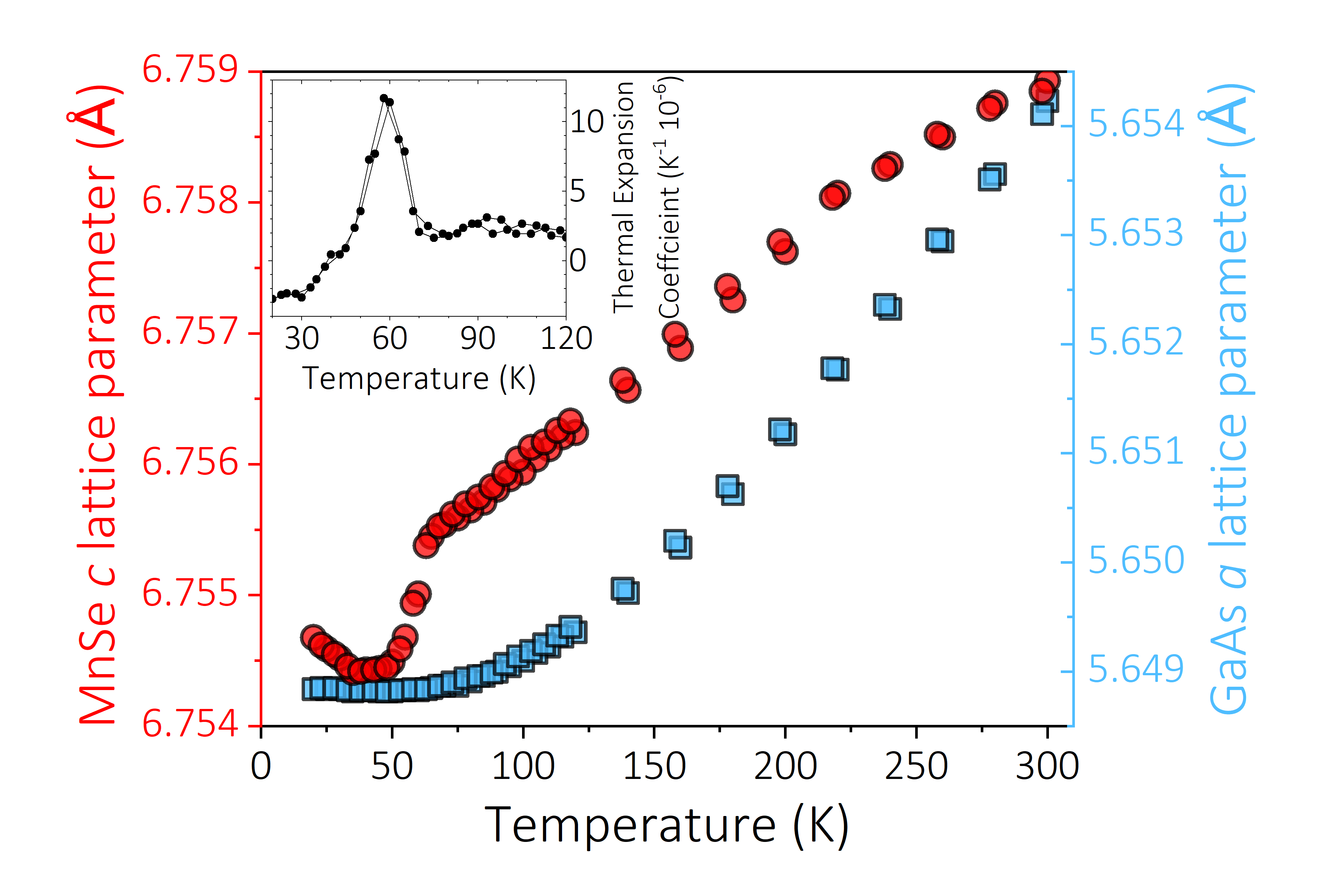}
  \caption{\label{fig:xrdtemp} Temperature evolution of the $c$ lattice parameter of MnSe (red circles) and $a$ lattice parameter of the substrate GaAs (blue squares) upon cooling and heating cycle. The inset shows thermal expansion coefficient of MnSe, which is proportional to the temperature derivative of the $c$ parameter, exhibiting a clear maximum at $T_N$.}
\end{figure}

Temperature-dependent XRD has been performed in symmetrical reflection geometry allowing for access to the out-of-plane lattice parameters. Temperature evolution of corresponding lattice parameters for GaAs and MnSe, as derived from refinements, is presented in Fig.~\ref{fig:xrdtemp}. The lattice parameter of GaAs decreases smoothly and monotonously upon cooling in a manner consistent with the Debye model of lattice expansion. At temperatures below 75 K the GaAs lattice parameter exhibits a plateau and becomes nearly temperature-independent suggesting harmonic regime of lattice vibrations. The behavior is fully reversible upon cooling and heating within experimental uncertainty. Previous studies of GaAs have reported weak negative thermal expansion in this temperature range \cite{Leszczynski:1997_APL}, resulting in a shallow minimum of the lattice parameter but such feature is not resolved in our measurements, possibly because its expected magnitude is comparable to the experimental sensitivity.

In contrast to GaAs, evidenced lattice changes for MnSe show a very different, anomalous behavior. Firstly, during cooling it hardly shows expected asymptotic approach to a linear function at temperatures above 200 K. Secondly, below $T=63\,$K a significant drop of the MnSe lattice parameter is evidenced. This anomaly, which is highlighted in the inset of Fig.~\ref{fig:xrdtemp}, shows the strongest effect on thermal expansion coefficient at 60 K. In context of further magnetic data, the anomalous behavior can be associated with the onset of long-range magnetic ordering in MnSe. This is a sign of strong magnetoelastic coupling similar to observed previously in case of a layered antiferromagnet CrPS$_4$ \cite{Fas:2025_AOM}. Thirdly, as apparent from Fig.~\ref{fig:xrdtemp}, different expansion of the substrate and the MnSe exert strain, that may influence superlattice properties. Lastly, for the MnSe at low temperatures some upturn of lattice parameter, below 32 K, is observed. Magnetic interactions may lead to hardening of the phonons due to additional forces that stiffen atomic bonds. This can be associated with an abrupt decrease of the lattice parameters at magnetic transitions. Actually, this is the case in MnSe at 62 K. However, the upturn at lower temperatures suggests that some interactions, presumably of magnetic origin, lead to softening of the lattice vibrations.   

\subsection{Magnetometry}

\begin{figure}
  \centering
  \includegraphics[width=1\linewidth]{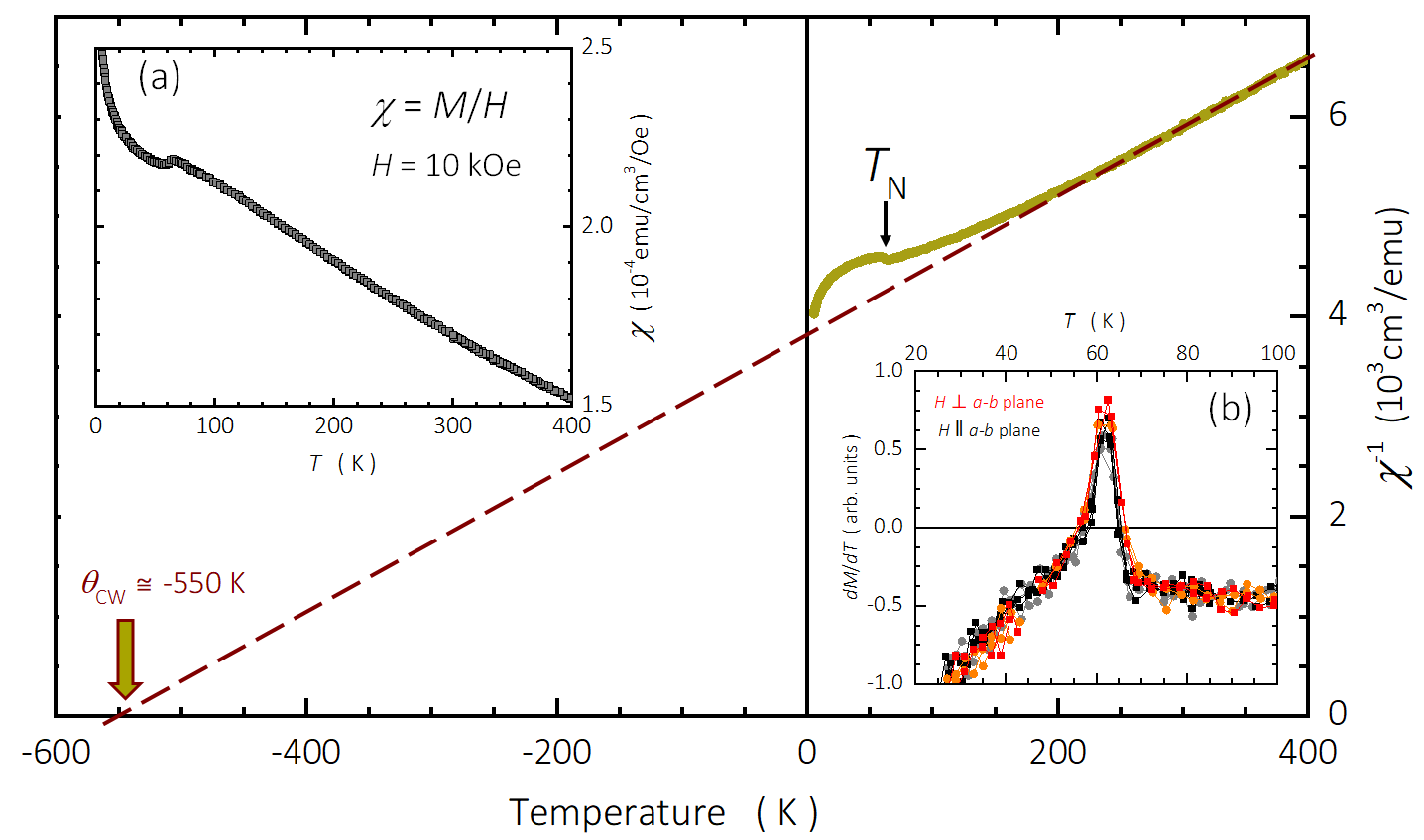}
  \caption{\label{fig:squid} Inverse susceptibility $\chi^{-1}$ as a function of temperature $T$ for the MnSe/CdSe superlattice sample. The left-hand side inset (a) shows susceptibility vs temperature whereas the right-hand side inset (b) depicts temperature derivative of magnetization $dM/dT$. Circles and squares depict measurements obtained at $H= 1$ and 10~kOe, respectively, whereas lighter and darker hues indicate warming and cooling, respectively. No matching factors are applied.}
\end{figure}

SQUID magnetometry confirms that the abrupt anomaly of the MnSe lattice parameter at $T=62\,$K corresponds to the onset of antiferromagnetic order in the wurtzite MnSe superlattice. The temperature dependence of the magnetic susceptibility, $\chi = M/H$, measured at $H = 10\,$kOe is shown in the top-left inset of Fig.~\ref{fig:squid}. The susceptibility exhibits a clear anomaly slightly above $60\,$K, marking the transition from the paramagnetic to the antiferromagnetic phase. The transition temperature is determined from the maximum in $dM/dT$, shown in the bottom-right inset of Fig.~\ref{fig:squid}. For both magnetic-field orientations with respect to the wurtzite $ab$ plane, and for both field values used in the experiment, $H = 1$ and $10\,$kOe, the derivative peaks at $T=62\pm1\,$K. We therefore assign $T_N = 62\pm1\,$K as the N\'eel temperature of the wurtzite MnSe superlattice.

Within the experimental resolution, no thermal hysteresis is observed between cooling and warming curves over the whole temperature range studied, distinguishing this behavior from that of the rock-salt phase of MnSe \cite{Grzybowski2024, Huang2019}. This indicates that the transition is consistent with a continuous paramagnetic-to-antiferromagnetic transition. The coincidence between the magnetic anomaly (Fig.~\ref{fig:squid}) and the abrupt change of the MnSe $c-$axis lattice parameter (Fig.~\ref{fig:xrdtemp}) provides independent evidence for magnetoelastic coupling at the onset of antiferromagnetic order.

To characterize the paramagnetic regime, we analyze the inverse susceptibility, $\chi^{-1}(T)$, shown in the main panel of Fig.~\ref{fig:squid}. Above approximately 250 K, $\chi^{-1}$ follows a linear Curie-Weiss behavior. This confirms that the high-temperature response is governed by localized Mn moments in the paramagnetic state. The Curie-Weiss fit yields an effective spin $S = 2.6 \pm 0.1$, close to the expected high-spin Mn$^{2+}$ value S = 5/2. This agreement indicates that the magnetic response involves essentially the full Mn sublattice of the wurtzite MnSe component.

The Curie-Weiss extrapolation gives a large negative Curie-Weiss temperature, $\theta_{CW} = -550 \pm 50\,$K, demonstrating dominant antiferromagnetic exchange interactions. The magnitude of $\theta_{CW}$ is almost an order of magnitude larger than $T_N$, which shows that the local antiferromagnetic exchange scale is much higher than the temperature at which long-range magnetic order is established. In the terminology introduced by Ramirez \cite{Ramirez:1994_AnnRev}, the ratio $f = |\theta_{CW}|/T_N$ provides a useful empirical measure of the suppression of the ordering temperature with respect to the exchange scale \cite{Ramirez:1994_AnnRev}. For the wurtzite MnSe superlattice investigated here $f$ is close to 9, a value indicative of a strongly reduced ordering efficiency.

The relevance of compositionally graded interfaces to the magnetic response was well established in II-VI diluted magnetic semiconductor superlattices \cite{Sawicki1994, Grieshaber1994, Kossacki:1995_SSC, Kossacki1999} as already indicated in TEM section. Indeed, the present system differs from an ideal bulk antiferromagnet in several important aspects. First, wurtzite MnSe phase is stabilized in a MnSe/CdSe superlattice rather than in a continuous bulk-like film. Second, the MnSe/CdSe interfaces are compositionally graded, so the Mn-rich regions are separated by intermixed (Mn,Cd)Se regions. Third, the wurtzite structure itself has recently been predicted to support several competing antiferromagnetic configurations with comparable energies, including collinear and non-collinear spin arrangements \cite{Mavani2026}. Therefore, the low value of $T_N$ relative to $\theta_{CW}$ likely reflects a combination of strong local antiferromagnetic exchange, finite-size and interface effects, and the presence of competing magnetic configurations allowed by the wurtzite lattice \cite{Samarth1991}. Therefore, the pronounced low-temperature upturn of $\chi(T)$, visible below about $30$-$40\,$K can be naturally explained by a Curie-like paramagnetic contribution from magnetically diluted Mn ions in the compositionally graded MnSe/CdSe interface regions. In these regions the local Mn concentration is reduced, so Mn ions have fewer Mn neighbors and are more weakly involved in the antiferromagnetic exchange network than Mn ions in the MnSe-rich cores. This contribution is weak at high temperature and therefore does not disturb the high-temperature Curie-Weiss analysis, but it becomes increasingly visible at low temperature. We therefore treat the low-temperature upturn (Fig.~\ref{fig:squid}(a)) as a diluted-Mn background superimposed on the antiferromagnetic response of the wurtzite MnSe superlattice.

The presence of antiferromagnetic correlations above $T_N$ is also supported by the shape of $\chi^{-1}(T)$. On cooling from high temperature, the experimental inverse susceptibility starts to deviate from the Curie-Weiss extrapolation well above the ordering temperature. Such a deviation indicates that short-range antiferromagnetic correlations develop in the nominally paramagnetic regime before long-range order is established. This behavior is consistent with the broad separation between the local exchange scale, represented by $\theta_{CW}$ and the actual ordering temperature, $T_N$. The observed magnetic anomaly at $62\,$K should therefore be understood as the temperature at which long-range antiferromagnetic order becomes established in the MnSe cores of the SL structure, whereas antiferromagnetic correlations are already present at higher temperatures.

The determined $T_N = 62\pm1\,$K is close to the recently established N\'eel temperature of wurtzite MnTe, $T_N = 58\,$K \cite{Sadowski2026}, and to the ordering temperatures reported for tetrahedrally coordinated zinc-blende Mn chalcogenides. This similarity suggests that the low ordering temperature is characteristic of tetrahedrally coordinated Mn-based chalcogenides \cite{Sawicki1994, Giebultowicz1993}. In contrast, the large negative $\theta_{CW}$ demonstrates that the exchange interactions in wurtzite MnSe are strong. The reduced ordering temperature therefore reflects the inefficient conversion of local antiferromagnetic exchange into long-range magnetic order.

The present result provides the first experimental determination of the N\'eel temperature of epitaxial wurtzite MnSe. Together with the Curie-Weiss analysis and the simultaneous anomaly in the MnSe lattice parameter, it establishes the wz-MnSe/wz-CdSe superlattice as an antiferromagnetic semiconductor with high-spin Mn$^{2+}$ moments, strong local antiferromagnetic exchange, and pronounced spin-lattice coupling. The proximity of competing magnetic configurations predicted for wurtzite Mn chalcogenides makes this system a promising platform for future microscopic studies of the magnetic ground state or element-specific magnetic probes.

\subsection{Optical properties}

\begin{figure}
  \centering
  \includegraphics[width=0.5\linewidth]{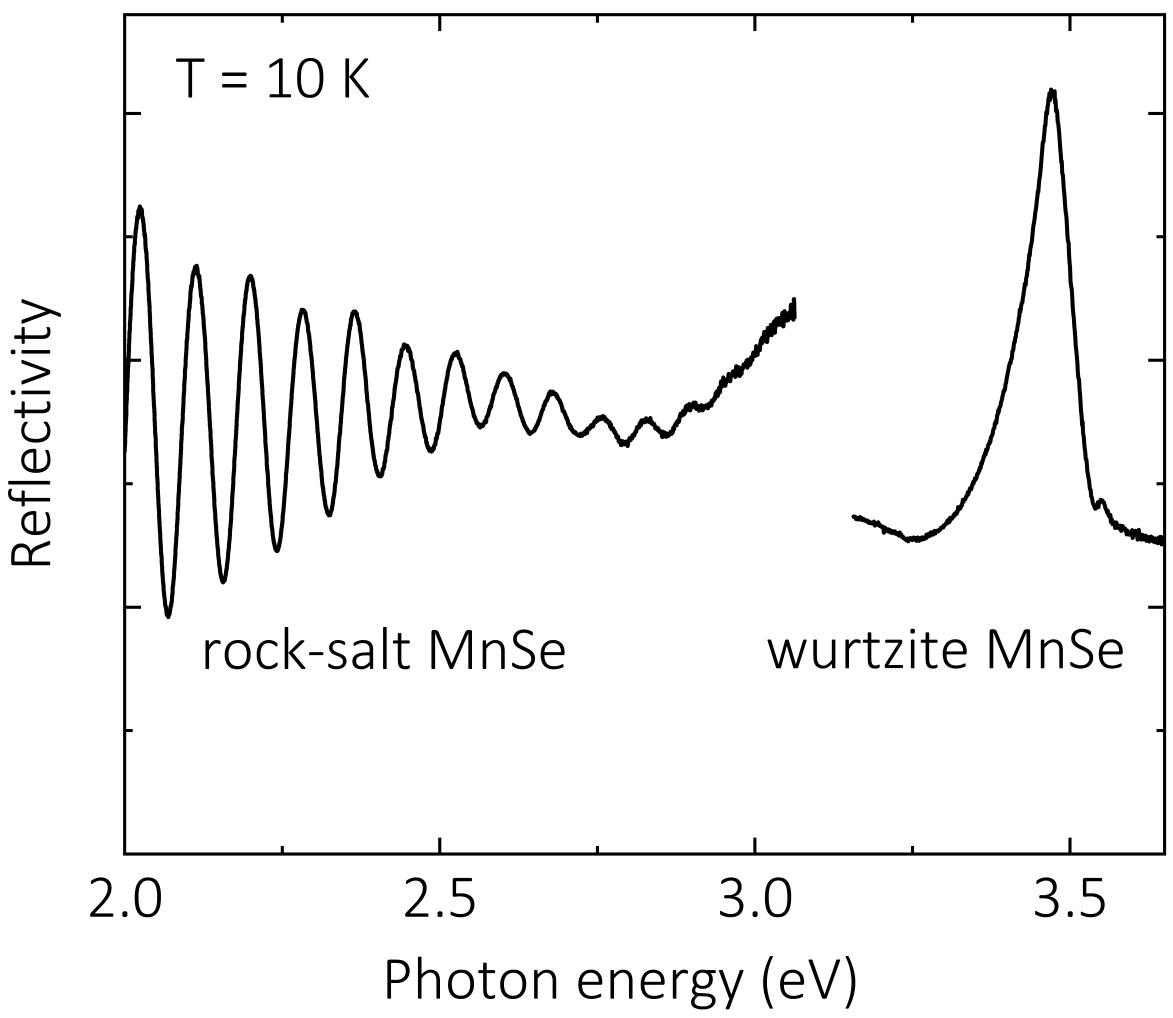}
  \caption{\label{fig:optics} Comparison of the reflectivity at cryogenic temperature $T=10\,$K for wurtzite MnSe/CdSe superlattice and a rock-salt MnSe $2.5\,\mu$m-thick layer.}
\end{figure}

The reflectivity spectra spectrum collected at $T=5\,$K in the spectral vicinity of the band gap for the wurtzite MnSe is shown in Fig.~\ref{fig:optics}). A signal from a reference, 2.5$\,\mu$m-thick rock-salt MnSe reference film in is also shown. A pronounced spectral feature is observed in the UV-range (3.4$\,$eV) for the wurtzite superlattice. We ascribe it to the excitonic transitions in wurtzite MnSe. Such spectral region of the band gap is in agreement with earlier studies of colloidal nanoparticles \cite{Sines:2010_AC}.  In contrast, rock-salt MnSe exhibits gradual decrease of the interference pattern amplitude around 2.8$\,$eV in agreement with earlier absorption and reflectivity studies \cite{Decker:1971_PRB, Kucharek:2019_APP}. The lack of sharp excitonic feature in the reflectivity spectrum of rock-salt MnSe is in agreement with the prediction of the indirect band-gap of this polytype, in contrast to expected direct band-gap in wurtzite form of MnSe \cite{Grzybowski2024}.

\section{Conclusions}

In conclusion, we achieved a large volume, wurtzite crystal phase of MnSe and determined its magnetic, structural and optical properties. A MnSe layers are stabilized within a micrometer-thick wz-MnSe/wz-CdSe superlattice. The total wurtzite MnSe thickness is two orders of magnitude larger than previously reported\cite{Grzybowski2024}. This overcomes a long-standing obstacle in accessing the wurtzite phase of MnSe, which otherwise reverts to its native rock-salt structure. The resulting periodically stacked wurtzite MnSe layers, exhibiting intriguing magnetic properties, are for the first time amenable to bulk-sensitive characterization methods enabling, for example, determination of the N\'eel temperature ($T=62\,$K) via SQUID magnetometry or observation of the excitonic transitions in optical measurements. Lattice parameters of wurtzite MnSe in the considered multilayered structure were determined to be $a = (4.18 \pm 0.03)\,$\AA\  and $c = (6.78 \pm 0.03)\,$\AA\ based on TEM and XRD measurement results\cite{uncertainty_note}. Our study highlights the ability to control and utilize the polymorphism of manganese chalcogenides up to micrometer-scale thickness through epitaxial growth engineering and opens the way for systematic studies of the relationship between crystal polymorphism, magnetic order, and the polar character of wurtzite MnSe.

\section{Methods}

MnSe/CdSe superlattice layers considered in this work are grown in a II–VI MBE growth chamber equipped with standard Knudsen effusion cells for both Se and Mn sources on GaAs (111)B substrate.

Transmission electron microscopy (TEM) investigations are performed using an image-corrected FEI Titan 80-300 Cubed transmission electron microscope operated at an accelerating voltage of 300~kV. For high-angle annular dark-field (HAADF) scanning transmission electron microscopy (STEM) imaging the Fischione 3000 detector is used. Cross-sectional TEM lamella is prepared using a focused ion beam FEI Helios Nanolab 600 system. Energy dispersive X-ray spectroscopy (EDX) data are collected in TEM by EDAX 30 mm$^2$ Si(Li) detector.

Room temperature XRD measurements are performed on an Empyrean X-ray diffractometer using Cu K$\alpha_1$ radiation $\lambda = 1.54056\,\text{\AA}$ with a hybrid two-bounce Ge(220) incident-beam monochromator. Low-resolution scans use a pixel detector; high-resolution scans used a three-bounce Ge(220) analyzer on the diffracted-beam side with a point detector.

Non-ambient X-ray diffraction (XRD) measurements were performed using MalvernPanalytical Empyrean diffractometer equipped with CuK$_{\alpha}$ radiation source. The PheniX low-temperature stage by Oxford Cryosystems, operating in between 20-300 K, was used. The stage was mounted onto motorized positioning system, that enables automatic calibration of the sample displacement during temperature change thanks to reference measurements of high purity tungsten (6N). The lines' profiles and positions were calibrated by measurements of the standard NIST660 sample. The resulting patterns were obtained in the $2\theta$ range of 20-100\textdegree  and refined by FullProf Suite package \cite{Carvajal2025}.

Magnetic measurements are carried out using an MPMS XL SQUID magnetometer (Quantum Design, San Diego, CA, USA) over the temperature range from 2 to 400 K. Measurements are performed with the magnetic field oriented either within or normal to the wurtzite $ab$ plane, corresponding to $H \perp c$ and $H \parallel c$, respectively. The contribution of the GaAs substrate is determined from an independently measured reference specimen prepared according to the same protocol. Its magnetic response is scaled by mass and subtracted from that of the MnSe/CdSe superlattice specimen. The resulting magnetic moment is normalized to the effective volume of the MnSe component. Careful determination of the substrate background and of the specimens’ geometry is required for quantitative SQUID magnetometry of epitaxial magnetic layers \cite{Sawicki:2011_SST, Ney:2006_JPCM,Stamenov:2006_RSI, Gas:2022_Materials, Gas:2026_PRApp}. To enable a direct comparison between the two field orientations, the specimen geometry is designed to remain effectively unchanged upon rotation, following the approach introduced in Ref.~\cite{Sztenkiel:2016_NC} and recently applied to wurtzite MnTe \cite{Sadowski2026}. All the measured samples are carefully cleaved into four strips approximately $6 \times 1.7$ mm$^2$ in size. The strips are stacked and bonded with diluted GE varnish to form a rod-like specimen with an approximately square cross-section. The resulting rod is then bonded, again using diluted GE varnish, transversely to a silicon strip used as the sample holder, with its long axis perpendicular to the holder axis and thus to the direction of the applied magnetic field. The two orthogonal measurement configurations are obtained by rotating the rod-like specimen by $90^\circ$ about its long axis while retaining the same transverse mounting geometry. This preserves the geometrical correction factor with respect to the SQUID pick-up coils for both configurations. Its value, $\gamma = 1.02$, is included in the magnetic data reduction.

A halogen lamp serves as a light source in reflectivity measurements. The beam is incident on the sample at a normal angle. The size of the spot on the sample surface is around 100 um. The spectra are registered by a Peltier-cooled CCD camera coupled to a 600 groves/mm grating spectrometer. The measurements are conducted at a temperature of 5 K.

\section*{Acknowledgements}

The research was funded in part by National Science Centre, Poland 2023/51/D/ST3/01978. The authors acknowledge the financial support provided by the Polish Ministry of Science and Higher Education through the SPUB subsidy for financing the maintenance of the Electron Microscopy and Nanolithography Laboratory (21/599012/SPUB/SP/2024). M. J. G. acknowledges the support within the IDUB project No. BOB-661-1196/2026 financed by the University of Warsaw.


%

\bibliography{MnSeCdSe_SL_acs_nano_template.bib}

\begin{mcitethebibliography}{64}
\bibitem[unc()]{uncertainty_note}
Weighted-average values were calculated neglecting correlations between the
  uncertainties.\relax
\mciteBstWouldAddEndPunctfalse
\mciteSetBstMidEndSepPunct{\mcitedefaultmidpunct}
{}{\mcitedefaultseppunct}\relax
\EndOfBibitem
\end{mcitethebibliography}

\end{document}